# Secure & Rapid composition of infrastructure services in the Cloud

Pierre de Leusse*, Panos Periorellis*, Paul Watson*, Andreas Maierhofer**
* School of Computing Science, Newcastle University
** Information Technology Futures Centre, British Telecommunications plc.
pierre.de-leusse@ncl.ac.uk

## Abstract

*A fundamental ambition of Grid and distributed systems is to be capable of sustaining evolution and allowing for adaptability [1, 2]. Furthermore, as the complexity and sophistication of theses structures increases, so does the need for adaptability of each component. One of the primary benefits of Service Oriented Architecture (SOA) is the ability to compose applications, processes or more complex services from other services which increases the capacity for adaptation. This document proposes a novel infrastructure composition model that aims at increasing the adaptability of the capabilities exposed through it by dynamically managing their non functional requirements.*

## 1. Introduction

In the past few years, inter-application integration has become one of the main interests in the IT industry [2-5]. This has brought up the emerging growth of the Service Oriented Architecture (SOA) paradigm which has become the main reference in terms of distributed software architectures [5]. Service orientation is a design paradigm intended for the creation of solution logic units that are individually shaped so that they can be collectively and repeatedly utilised in support of the realisation of a specific set of strategic goals and benefits associated with SOA and service-oriented computing [6].

In such environments where change can be frequent, adaptation to contextual changes is a strong requirement [7] but, due to its complexity, can be difficult to achieve. Indeed, the SOA maturity model described in [8] defines the ability to automatically react and respond to change as one of the main characteristics of its top level environments. Such changes can come from the need to adapt the non-functional behaviour (e.g. Security, QoS...) of a service according to different contexts.

Service oriented computing and its proposed methodology for designing services that are operating as autonomously as possible, has also given rise to alternative ways of thinking regarding application design and more importantly application delivery. S3 [9] and EC2 [10] services that Amazon offers demonstrate clearly the power of SOA from a developer's point of view, but also highlight the power gained by the end user, who can combine services on demand to create his custom browser based application. The notion which is now becoming widely addressed as Cloud computing characterises such flexible systems. In this paper we acknowledge the new trend of Cloud computing and we are bringing forward a new set of requirements –as our example demonstrates- within which an execution environment is created in which application services can offload non functional requirements to contextualised on demand services provisioned from the cloud. In our case, the parameters governing the execution environment are defined at context creation time, and can be adopted at any time in the execution process by authorised entities. This approach has several advantages for both application providers and users.

In a first part, the virtual music store scenario will be presented as a use case. Following this, the architecture of the intended solution will be introduced. Finally the current state of the work as well as its future will be discussed.

## 2. Virtual music store Scenario

### 2.1. Description

This section describes the virtual music store as an example of a Virtualised Organisation (VO). VOs can be loosely defined as temporal collaborations between organisational entities [11]. According to [12], VOs are frequently restructured, sustained to capture the value of a market opportunity and dissolved again to give way for the creation of a next VO from within the network of independent partners. This represents a need for adaptability that current systems, such as the GOLD middleware [13] or the current B2B gateway [14, 15], attempt to address by providing one type of security profile.

The aggregated services are virtual music stores serving specialised markets or communities of interest. The basic service providers include copyright owners of musical recordings or their representatives who make these recordings available online and syndicated blogs or review sites. The music stores reach agreement with music providers enabling them to act as re-sellers of bundles of recordings from their catalogues. The virtual store is a VO consisting of the music store operator as well as content providers and it runs on top of an infrastructure provided by an infrastructure provider.

The end customer of the virtual music store will be a member of the public. What they will see is a normal web-site where they will be able to search for and buy tracks and read reviews and blogs. This could be presented to them in much the same way as AbeBooks does, i.e. a search page and then each returned item is linked in from an independent seller; or stores could hide the aggregated nature of the service.

## 2.2. Partners and Roles

In the music store, the main categories of partner are:

- Infrastructure provider

This role involves providing the VMS with a Virtual Hosting Environment (VHE), the B2B gateway. The purpose of the infrastructure is to hide the technical complexity of the middleware involved to the different participants in the virtual music store.

- Music content provider

This is specialist content provider (eg. record labels or other copyright owners).

- Virtual music store operator

The broker of music. It is assumed that the store operator will be the VO Initiator. As such the operator is responsible for instigating the opening federation process.

- Value adding service provider

This is a third party entrusted with providing Value Adding Services (VAS). These services provide non functional proprieties (eg. Security, audit, translation) and allow the content providers and music store operator to leverage on the VHE to enhance their interoperability and quality of service.

## 2.3. VO Lifecycle

This section describes the life-cycle of the virtual music store. The music store life cycle starts with the initial agreements and discovery of the potential partners, the VO foundation. This is followed by the negotiation between these partners and the VO initiator to reach a firm collaboration agreement, this stage is called the partners' federation. Following this, the capabilities are virtualised and made available to the partners in the newly formed VO. Finally the adaptability faculty of the virtual store infrastructure is introduced.

**2.3.1. VO foundation.** Prior to any task and once the virtual shop has decided to establish the music store, it needs to reach an agreement with the infrastructure provider. The infrastructure provider is said to provide a VHE, on which it instantiates an 'empty' VO which is configured by the virtual shop operator.

The VHE being in place, the shop operator contacts the potential participants of the music shop (i.e. content providers). Agreements are reached between these music providers and the shop operator and access to the VO Manager is granted to the content providers to setup their accounts and modify their data.

The content providers can consequently publish the business functions they want to expose. These selected capacities are published as services into a capability registry.

**2.3.2. Partner federation.** At this stage it becomes possible for the virtual shop operator to put in place the different services offered by the music store.

To achieve this, as introduced above, the operator creates a new VO for each federation of content providers it wants to create. Additionally, the operator defines a collaborative process to describe the interactions between the different business functions potentially present in the federation. Once this structure is in place, the operator can search the capability registry for the specific business functions it wants to aggregate and using the VO Manager sends a participation request to the relevant providers.

The providers contacted can inspect the process description and interaction description already provided by the store operator to take a decision upon participating in this federation. Having accepted the invitation, the content providers associate to the VO the VAS profile they want to apply to this federation. The VAS profiles are defined for each capability by its provider.

These profiles include infrastructure services used to secure and monitor the services. They are created and managed in much the same way as the federation between the operator and the music providers but include the VAS providers. The services are typically comprised of policy enforcement, authentication, authorisation and other added value services such as billing or auditing. In addition, the profiles are composed of policy templates that define the policies to be applied to each of the selected infrastructure services in the profile.

Following this, the operator can review and select the best matches in the positive answers it has received. With all the targeted business functions fulfilled, the virtual shop operator can continue the VO creation process and sends a creation order.

The VO management tool subsequently interacts with each partner's gateway. To allow the different identity providers to recognise each other's authority, it sends the relevant list of business cards associated with each business partner (role). In addition, the VO management tool sends the policies related to the implementation of the collaboration management for each business function. These policies come on the top of the security profiles setup and made available by the service providers.

Brokered services, such as jazz music store aggregating the different content providers' services that offer jazz music, can be created along with their federation data following this method.

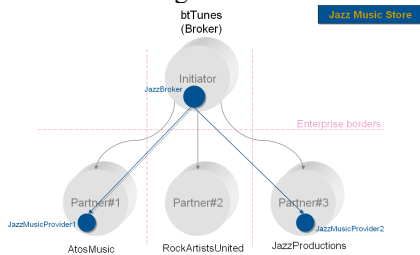

**Figure 1. The Jazz Music Store VO**

**2.3.3. Capability virtualization.** With the federation in place, it is possible for the participants to finalise the configuration of the instances of the services they selected and prepared for this specific VO. Before undertaking this, the gateway management interface allows the participant to inspect the configuration of its infrastructure. At this stage, the configuration of the infrastructure will have evolved as the services are exposed and activated. Additionally, the selected Federated Identity Provider (FIP) has built trust with the FIPs of the other partners in the VO. Furthermore, the baseline policies that restrict who can issue access policies about which resources have been activated. Finally, the VAS profile that will be applied by the Partner for the business functions it performs is stored in a specific registry. To keep track of the configuration, the settings are associated with a unique collaboration ID.

Upon configuration of the infrastructure, provisioning of policy templates and establishment of trust between the different VOs it becomes possible for the capability instances exposed to be invoked within the context of the virtual music store.

**2.3.4. Adaptability.** A music provider might want to participate in several such federations to increase its visibility. But different partners in various VOs will have distinct security needs and settings. By adjusting the VAS profile used in each federation to its specific needs, the content provider can more promptly offer its services.

# 3. Architecture

The infrastructure model presented in the following chapter is inspired by autonomous/adaptive computing architecture such as the Self-Managed Cell (SMC) presented in [16]. The objective of these architectures [17, 18] is to promote self-configuration, self-healing, self-optimisation and self-protection. Although the model described below does not fully reach these goals, mostly for security reason, its intention is to provide adaptability for resources and therefore adopts the same ambition.

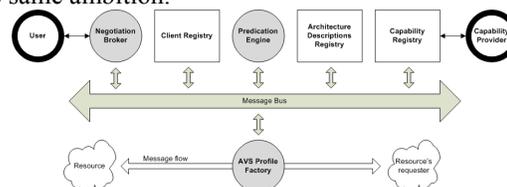

**Figure 2. Secured Profile Management System**

## 3.1. Negotiation Broker

The client broker is the interface that allows resource owners to define their requirements in term of the VASs added on the messages that come from or towards their resources. The request consists of a list of services required by the owner along with the operation type the VAS profile is required for (e.g. request or response). The request is expressed using an ontology that is provided by the Profile Management system. It consists of a list of components with potential constraints attached to them. These constraints could include QoS (e.g. throughput, answer time) details for the components used as well as specific semantics to be used for certain operations (e.g. XACML [19], SecPAL [20]).

Alternatively or additionally, the resource owners can declare adaptability level constraints. This could be expressed in such a manner that all requests coming for certain partners in specific federations should be accommodated as best as possible. Or at the opposite that the choices made in terms of VASs required should not be tempered with, or require the client's authorisation to go further.

The second role of the Negotiation Broker is to allow other infrastructures such as the Profile Management system to communicate and potentially negotiate terms of the VAS, in order to allow for compatibility between the resource's requester and the resource's interface as exposed for this particular usage.

For instance, in the virtual music shop scenario described above, the virtual shop operator would send a participation request to a music provider. Upon its reception, the content provider would check the collaborative process to see how its content would be used and could decide to define a very open adaptability level, trusting the VO initiator in its security choice. Alternatively, the music provider seeing that competitors could be able to access its pricing policy could require a specific VAS profile.

## 3.2. Client Registry

The client registry holds data related to a particular user. This data includes the different requests for VAS Profile as well as adaptability level constraints, the different composition models accepted, both abstract and concrete, as well as the degree of acceptance shown in front of potential alternatives proposed by the Predication Engine. Finally, usage frequency, failure rate and different critical data related to a particular user and profile could also be held in this registry.

## 3.3. Predication Engine

The role of the predication engine is to determinate the best possible way to achieve the VAS profile requested. The decision making process is based on the requirements given by the user, the capabilities held by the system along with their associated constraints and the architectural models stored in their registry.

In order to achieve this, the engine goes through a first stage that aims to create an Abstract Generic Composition Model (AGCM). The AGCM validates the user's request against an ADM, this one being given nominally by the user or assigned dynamically by the system. During this stage, the VAS profile is being checked for completeness (e.g. missing mandatory components) and the different VAS are being ordered. The AGCM is expressed using the same semantics as the request and defines the SP in terms of abstract components needed.

Once the closest component based AGCM possible to the user requirements has been built, the matching capabilities are selected and their compatibilities checked. The system processes the AGCM, and for each component described, attempts to find a capability or another AGCM that completely realises it. This allows the creation the Abstract Specific Composition Model (ASCM) which is a location specific version of the AGCM. This stage, along with the previous one, corresponds to the planning sub-cycle as defined by the ASG Semantic Enterprise Platform [21].

With the ASCM completed, the concrete capabilities can now be searched and once found, bound the Concrete Composition Model (CCM). This discovery and selection could be made using non functional properties. This stage corresponds to the binding sub-cycle as defined by the ASG Platform [21]. This last stage results in the creation of a CCM which, depending on the agreement can be proposed to the user.

Once the CCM has been validated, the required capabilities can be instantiated; the Secured Profile (SP) can be deployed and made ready to be used. This stage corresponds to the enactment sub-cycle as defined by [21].

If the user doesn't accept the proposed CCM, another cycle is started with different requirements.

Note that in the future, the Predication Engine could make different suggestions (CCMs) by itself based on the knowledge of the user's requirements and level of adaptability as well as the capabilities' constraints.

## 3.4. Architecture Descriptions Registry

In this registry the descriptions of the different requirements and constraints for the allowed architectures are kept. These descriptions are expressed using the same formal model as the one applied to validate the profile. The registry could be organised in categories with such architectures as Liberty Alliance in the Authentication category.

## 3.5. Capability Registry

The VAS registry allows the system to hold the location, description and constraints related to the VAS. In addition to these traditional elements, the Capability Registry will hold information about how to set up the VAS. For instance, a Policy Decision Point (PDP) might need to build trust with a Security Token Service (STS) prior to any interaction. This process as well as the VAS descriptions should be defined and expressed using the same ontology as the one used for the client's request.

### 3.6. Secured Profile (SP) Factory

This component is divided in two different parts, the SP enactment engine and the message broker, together they allow for the VASs to be put in place. The enactment engine retrieves the relevant CCM from the Client Registry at run time and implements the appropriate SP. The message broker intercepts the messages from and towards the resource, contacts the enactment engine and interacts with it before forwarding the message as appropriate.

### 3.7. Lifecycle Management

Once the system has established that the Profile is valid and recognised a way to enact it, it becomes necessary to handle its lifecycle. This is achieved thanks to the agreement reached between the different partners (e.g. client, Infrastructure Provider…) on service availability and potentially QoS. According to this agreement, the VAS Profile can be made available and managed in several different ways. First, the profile is set up and exposed as soon as possible after the request for it has been made and kept available as often as possible. The second option is to set it up and expose it only at specific times in conformity with the agreement mentioned previously. Finally, the VAS Profile is made available only when it is needed. These are the three main options but variations between the times the different stages of validations and enactment are made are possible.

Internally, it is be possible for the secured gateway to store the profile at different stages of validation or enactment in the Client Registry to improve the performances of the process of VAS Profile enactment.

### 3.8. Adaptation

In a federated model such as the virtual music shop scenario, the different partners must be able to synchronise their common processes and semantics (e.g. authentication and authentication protocols). The initiator of the synchronisation would be the federation initiator or the member needing a change. In order to allow this, standardised interfaces will be available for the partners to communicate to each other. This will require the potential building of trust before the full completion of the profile's enactment. Alternatively customised forms are offered for participants to express such requirements during the formation of the VO. The token is then sent by the requestor to the resource owner for validation. Upon validation access to the resource is granted.

As it is mentioned one of our aims is to alleviate any constraints regarding message composition from the users. As such the method can be used to retrieve such requirements from the users service interface. Such requirements can be expressed in an ad hoc manner or make use of standards such as WS-Policy [22] and WS-Policy Attachment specifications [23] that specify ways of expressing such requirements and attaching them on WSDL interface descriptions. So effectively, composition requirements of a message can be expressed in a WS Policy document and attached by reference to a WSDL description document. Although WS Policy is an extensible standard we acknowledge the difficulty in setting up semantics for a particular domain. The fore mentioned WS-Policy Attachment specification provides a standard schema for expressing security related requirements and a protocol for exchanging information to obtain such requirements.

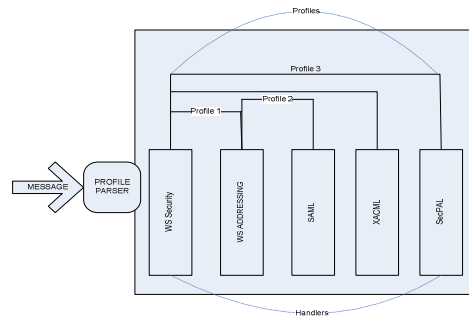

**Figure 4. Dynamic profiling**

As figure 4 suggests end users can express their profile requirements in via some protocol which inform interested parties of their preferences and consequently trigger a profile creation at the recipient's end. In the above figure we suggest that by doing so we can enable security related profiles to be created on the fly or on demand as services are discovered. A subset of the available handlers is utilised in order to accommodate or enable communication between 2 parties. Several profiles can be created depending on the type of user. Although it is not shown in the figure the profiling infrastructure can be independent of the service itself alleviating the service from this extra workload.

Furthermore, in addition to the actual adaptation strategy (e.g. modify access right), refinement of the profile and perhaps change of the profile's architecture could be made in some cases (e.g. specific types of attacks might require different security architectures, authentication mechanism…). This would have to be made in accordance to the user's adaptability level constraints and potential validation.

## 4. Conclusions and future work

In the paper we have introduced an infrastructure model that aims at increasing the adaptability and potentially the security of the resources exposed through it. This is achieved by a mediator exposing a virtualised interface of the resource enhanced with a SP. The SP itself will be formed using semantic technologies linked with a formal model. The semantic technologies will describe the different components along with their constraints and allow for their dynamic selection and composition. The formal model behind it will permit to insure the viability of the composition or SP as well as allowing improving it on the fly.

The current state of this project is the BT Secured B2B GW [14, 15]. It allows an enterprise to expose different capabilities as web services in a secure, dynamic, and virtualised manner. The virtualisation is guaranteed via the creation and management of service instances which contain infrastructure-specific configuration including security parameters.

The next stage in this project is to define the ontology describing the components as well as the formal model [24]. We anticipate that the current state of the research in both domains will allow the quick development of a working prototype.

There are a number of issues still to be resolved, such as making sure the communication protocols between the different B2B GWs or Secured Profile Management System allow for smooth negotiations. All the details of the different components and the way they communicate to each other are also still to be concretely defined.